Paper Title:

# A Business Maturity Model of Software Product Line Engineering

**Authors:**

1. Faheem Ahmed, Ph.D.

   College of Information Technology, PO Box 17551, UAE University, Al Ain, UAE
   f.ahmed@uaeu.ac.ae

   Tel: 00971-50-935-7086, Fax: 00971-3-767-2018

2. Luiz Fernando Capretz, Ph.D.

   Department of Electrical & Computer Engineering
   University of Western Ontario, London Ontario, Canada, N6A 5B9
   lcapretz@eng.uwo.ca

   Tel: 1-(519) 661-2111 Ext (85482), Fax: 1-519-850-2436

**Abstract:** In the recent past, software product line engineering has become one of the most promising practices in software industry with the potential to substantially increase the software development productivity. Software product line engineering approach spans the dimensions of business, architecture, software engineering process and organization. The increasing popularity of software product line engineering in the software industry necessitates a process maturity evaluation methodology. Accordingly, this paper presents a business maturity model of software product line, which is a methodology to evaluate the current maturity of the business dimension of a software product line in an organization. This model examines the coordination between product line engineering and the business aspects of software product line. It evaluates the maturity of the business dimension of software product line as a function of how a set of business practices are aligned with product line engineering in an organization. Using the model presented in this paper, we conducted two case studies and reported the assessment results. This research contributes towards establishing a comprehensive and unified strategy for a process maturity evaluation of software product lines.

Key Words

Software Product Line
Software Process Assessment
Maturity Evaluation
Business Process
Organizational Management
Software Process Model

# I. INTRODUCTION

One of the major concerns of software development organizations is the effective utilization of software assets, thus reducing considerably, the development time and cost of software products. A significant number of organizations which trade in wide areas of operation, from consumer electronics, telecommunications, and avionics to information technology, are using software product lines practice as it effectively makes use of software assets. Clements *et al.* (2005) report that software product line engineering is a growing software engineering sub-discipline, and many organizations including Philips®, Hewlett-Packard®, Nokia®, Raytheon®, and Cummins® are using it to achieve extraordinary gains in productivity, time to market, and product quality. Clements (2001) defines the term "software product line" as a set of software-intensive systems sharing a common, managed set of features that satisfy the specific needs of a particular market segment or mission, and are developed from a common set of core assets, in a prescribed way. Some alternative terminologies for "software product line" that have been widely used in Europe are; "product families;" "product population;" and "system families".

The acronym BAPO (*van der* Linden, 2002) (Business-Architecture-Process-Organization) defines the process concerns associated with software product lines. The "Business" in BAPO is considered critical as it deals with the way in which the products resulting from a software product line make profits. Software is perhaps the most crucial piece of the business entity in this modern marketplace, where important decisions need to be made rapidly. The organizations that fail to respond with sufficient speed have generally lower chances of survive. Business is perhaps the most crucial factor in the software product lines, mainly due to the necessities of long-term strategic planning, initial investment, payback period and retaining the market presence. Business requires continuous monitoring and evaluating of customers, competitors, market segment, marketing strategies, financial, assets management, etc. The business dimension of software product lines deals with managing a strong coordination between product line engineering and the business aspects of product line. Business assessment is essential for improving the overall software product line engineering process because it provides information about the maturity of an organization in doing the business of software product lines and also highlights the areas that require improvements. This paper presents a comprehensive methodology for the business assessment of organizations dealing with software product lines, thus addressing a topic of immense importance from the perspective of software engineering economics.

## A. Software Product Line Engineering Maturity Model: The Big Picture

The maturity assessment of software process within an organization has always been a key research area in software engineering. The Capability Maturity Model (CMM) proposed by the Software Engineering Institute (SEI) has been accepted as the de facto standard by the software industry. According to Paulk *et al.* (1993) CMM provides software organizations with guidance on how to gain control of their processes for developing and maintaining software and how to move toward a culture of software engineering and management excellence. The objectives of CMM are to provide a guideline to software development organizations to determine the current process maturity and to develop a strategy for improving software quality and process. CMM further evolved into CMMI. Jones and Soule (2002) discuss the relationship between software product line process and CMMI model and observed that software engineering process discipline as specified in CMMI models provides an important foundation for software product line practice. Jones and Soule (2002) concluded that in addition to the key process areas of the CMMI model, software product line requires mastery of many other essential practice areas. Although they have compared the process areas of software product line and CMMI and found some similarities, they emphasized there is a need to establish a comprehensive strategy for process assessment of software product line in particular, which requires identification of those *Key* Process Areas, which are not currently a part of CMMI. SEI proposed Product Line Technical Probe (PLTP), which is intended to assess an organization's ability to adopt and succeed with the software product line approach. PLTP is based on framework for software product line practice (Clements, Northrop 2002). Within PLTP there are 29 practice areas, which are divided into three categories; product development; core asset development; and management. The framework does not clearly define any levels to be assigned to an organization in order to know the maturity of the current process but rather simply identifies those potential areas of concern that should be given attention while carrying out any software product line activity.

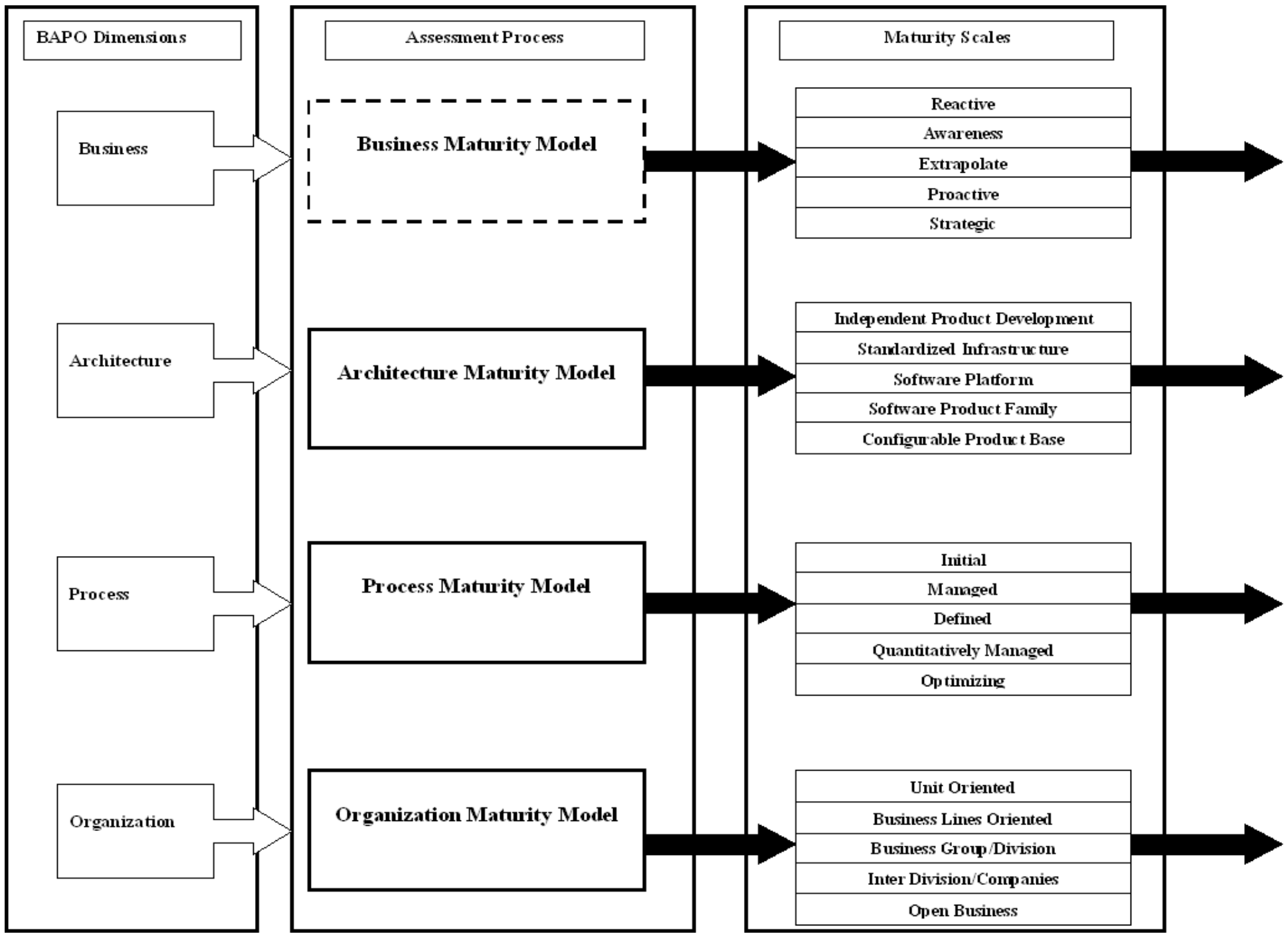

**Figure 1: Software Product Line Engineering Maturity Model: The Big Picture**

Software process maturity evaluation has been a key research area in the software research community because of its impact on the productivity of the development process. Software product line is a relatively a new concept in the history of software development and business. A lot of effort has been spent on the process methodology and the industrialization of this paradigm. The organizations dealing with software product lines also require a methodology to evaluate the maturity of software product line process. *van der* Linden *et al.*(2004) propose a four-dimensional software product line maturity evaluation framework based on the BAPO concept of operations. It provides a foundation for systematic and a comprehensive strategy for the process maturity evaluation of software product line. Figure 1 illustrates the conceptual layout of this maturity evaluation approach. The four dimensions of the framework are: Business, Architecture, Process and Organization. *van der* Linden *et al.* (2004) identified a maturity scale of up to five levels in ascending order for each dimension of BAPO, as shown in the rectangle of "Maturity Scales" in Figure 1. In the case of software product lines, this results in separate values for each of the four dimensions. *van der* Linden *et al.* (2004) proposed that the "P" which is "process" in BAPO is the software engineering process whose maturity can be found out by using any one of the popular software engineering process assessment approaches such as CMM, BOOTSTRAP or SPICE etc. The maturity models for other dimensions of business, architecture, and organization have not as yet been given a great deal of attention by software engineering community.

This work presents a business maturity model for software product line. The model provides a methodology to evaluate the current maturity of the software product line business of an organization. This is the first study of its kind within the area of software product lines to the best of our knowledge. It is important to note here that, the maturity models to evaluate the other two dimensions of BAPO are beyond the scope of this study as this work concentrates on only the business dimension. The dashed rectangle in Figure 1 clearly highlights the scope of the work presented in this paper in the overall software product line maturity evaluation process. The main objective of this research is to contribute towards a unified strategy for process evaluation of software product lines.

## B. Software Product Line Business Dimension: Related Work

Bayer *et al.* (1999) at the Fraunhofer Institute of Experimental Software Engineering (IESE) developed a methodology (PuLSE, Product Line Software Engineering) for the purpose of enabling the conception and deployment of software product lines within a large variety of enterprise contexts PuLSE-Eco is a part of Pulse methodology that deals with defining the scope of software product lines in terms of business factor. PuLSE-Eco identifies various activities, which directly address the business needs of software product lines such as system information, stakeholder information, business objectives and benefit analysis. *van der* Linden *et al.* (2004) identified some main factors in evaluating the business dimension of software product line such as identity, vision, objectives and strategic planning. They classified the business maturity of software product line into five levels in ascending order: reactive, awareness, extrapolate, proactive and strategic. Clements and Northrop (2002) highlight customer interface management, market analysis, funding, and business case engineering as important activities from the perspectives of organizational management. Kang *et al.* (2002) present a marketing plan for software product lines that includes market analysis and marketing strategy. The market analysis covers need analysis, user profiling, business opportunity, time to market and product pricing. The marketing strategy discusses product delivery methods. Toft *et al.* (2000) propose the "owen molecule model" which consists of three dimensions of social, technology and business. The business dimension deals with setting up business goals, and analyzing commercial environment. Fritsch and Hahn (2004) introduce Product Line Potential Analysis (PLPA) which is intended to examine the product line potential of a business unit through discussions with managers of the business unit, because in their opinion they know the market requirements, product information and business goals of the organization. Schmid and Verlage (2002) discuss a successful case study of setting up software product line at Market Maker and highlights market and competitor analysis, vision of potential market segment and products, as significantly important activities. Ebert and Smouts (2003) weight marketing as one of the major external success factors of product line approach and further concluded that forecasting, the methods used to influence the market, a strong coordination between marketing and engineering activities, are required for gaining benefits from product line approach.

The summary of the related work presented in this sub-section highlights some key business practices such as strategic planning, innovation, market orientation, business vision, order of entry, financial management and customer orientation. Other literature survey's (Aguilar-Sav'en, 2004; Bergstrom, 2000; Davenport, 1993; De Castro and Chrisman, 1995; Wappler, 2000; Verlage and Kiesgen, 2005; Weiss and Lai, 1999) also pointed out the significance of these key business factors in the overall business performance of an organization. In order to examine the effect of these key business factors on the overall performance of software product line engineering process in an organization Ahmed and Capretz (2007) conducted a quantitative survey of software organizations currently involved in the business of developing software product lines over a wide range of operations, including consumer electronics, telecommunications, avionics, and information technology. The main objective of this study was to identify the effect of business factors in the performance of software product line and to provide a rationale to the structure of the business maturity model presented in this paper. The results of the study provide evidence that organizations in the business of software product line development have to manage multiple key business factors to improve the overall performance of the business, in addition to their efforts in software development. With empirical evidence that these factors contributes in the business performance of an organization managing software product line engineering we used these key business practices as the foundation of the maturity model presented in this paper to evaluate the business maturity of software product line of an organization.

# II. THE BUSINESS MATURITY MODEL OF SOFTWARE PRODUCT LINE ENGINEERING

The business maturity model of software product lines aims to establish a comprehensive strategy to evaluate the business dimension of software product line process. It describes the business assessment methodology of software product lines, determines the current maturity of software product line business of an organization and identifies its strengths and weaknesses. It is structured in a way to determine how various business practices are carried out and resources allocated to software product lines development. The business dimension of software product line does not only take net cash flow in and out as parameters to evaluate the maturity of business. Rather, it assumes a strong coordination between product line engineering and the business aspect of product line and evaluates the maturity of the business as a function of how set of business practices are aligns with product line engineering. The functional structure of the model consists of a set of questionnaires purposely designed for evaluating the maturity at each of the five maturity levels in ascending order of reactive, awareness, extrapolate, proactive and strategic. A survey of work carried out in the business dimension of software product lines reported in Sub-Section "B" of Section I along with a survey in business and management theories provides the foundations for designing the questionnaires.

**Table-I: Configuration of Business Maturity Model**

| Dimension No. | Business Dimension | Practice No. | Business Practice |
|---|---|---|---|
| 1 | Marketing strategy | 1 | Market orientation |
| | | 2 | Relationships management |
| | | 3 | Order of entry to the market |
| 2 | Portfolio management | 4 | Financial management |
| | | 5 | Assets management |
| 3 | Business planning | 6 | Strategic planning |
| | | 7 | Business vision |
| | | 8 | Innovation |

## A. Configuration of Business Maturity Model

The functional configuration of the business maturity model for software product lines consists of a set of three business dimensions: marketing strategy, portfolio management and business planning, and eight business practices spread over those dimensions. Table-I defined the hierarchy and domains of the business maturity model for software product lines. The marketing strategy dimension covers the practices of market orientation, relationships management and order of entry to the market. Portfolio management deals with practices of financial management and asset management. Business planning spans strategic planning, business vision and innovation. In this paper we refer the term "Business Practice" as the activity that in conjunction with software product line engineering contributes to the business performance of an organization. The term "Business Dimension" refers to a set of interrelated business practices, which cover the business dimension of product line engineering. These business practices lay the foundations for the set of questionnaires, which have a number of questions about how effectively these practices are performed in the business dimension of software product lines. The subsequent sub-sections elaborate the concepts these business practices employ in detail.

### 1) Marketing Strategy: Literature Review

The marketing strategy dimension covers market orientation, relationship management and order of entry to the market. The concept of market orientation provides an advantage over the competitors by identifying what customers want and by offering products that are different and superior to those offered by competitors. Market orientation deals with the acquisition, sharing, interpretation and utilization of information about customers and competitors. According to Kohli and Jaworski (1990) market orientation is the generation or acquisition of market intelligence pertaining to current and future customer needs. Conversely, Narver and Slater (1990) consider market orientation as an organizational culture that most effectively and efficiently creates the necessary behaviors for the creation of superior value for buyers. The market orientation consists of three dimensions: customer orientation, competitor orientation, and inter-functional coordination. Software product line requires an in-depth knowledge of the market, which helps in capturing requirements of product line. Birk *et al.* (2003) define market orientation in context of software product lines as whether the organization targets a specific market segment without a specific customer in mind or addresses individual customer projects. The software product line deals with developing a considerable number of products to capture various market segments, thus providing justification for a product line. Market orientation provides imperative information about the concerns and requirements of customers, which needs to be accommodated in the successive products from a product line. PuLSE-Eco (Knauber *et al.*, 2000) illustrates various activities associated with market orientation for successful adoption of software product lines concept in an organization. It considers collecting and analyzing stakeholders' information as helpful in defining the product line scope.

Wilson (1995) observes that relationship management is concerned with the development and maintenance of close, long-term, mutually beneficial, and satisfying relationships between individuals or organizations. Crosby *et al.* (1990) consider relationship management as the extent to which parties have the orientation or behavioral tendency to actively cultivate and maintain close working relationships. The organizations that have established close relationships with their customers are generally more successful in maintaining profitable businesses. Some factors which contributes to the development of good relationship management are the management of customer information, customer profiling, customer support and services, promotional strategies, channel management, and organizational behavior. Business success is highly dependent on the extent to which customers are satisfied with product and services of an organization, as well as how they establish the loyalty of customers by improving their relationship management. Software product line entails the development of multiple products within a common application domain. Excellent relationship management builds a mutual confidence between customers and the firm, thus allowing the organization to convince customers about the new products.

The appropriate time for technology-based products to enter into the market is even more critical for the profitability and competitive position of an organization. The right product at an appropriate time of launch has a higher potential of success. There are three observed categories in a firm's order of entry in the market: pioneers, early followers, and late movers (Ansoff and Stewart, 1967; Robinson et al., 1992). The benefits of being the first in the market have long been recognized in the business sector; and pioneers can gain a sustainable competitive advantage over followers because initially, they are the only solution providers in the market segment and subsequently capture a major portion of that market. Order of entry is perceived as a crucial business decision, which has a long lasting profound impact on the performance of an organization in capturing and retaining the market. Software product line has the potential of capturing market by introducing new products. Appropriate timing to launch a new product out of product line is critical. The order of entry depicts the delivery schedule and helps in setting up the production plans of product line engineering.

## 2) Portfolio Management: Literature Review

Portfolio management covers the financial and asset management. Financial management deals with making decisions about fiscal matters within an organization. A financially strong organization envisions business progress, especially in terms of income, balance and cash flow. Effective financial policies lead to successful businesses. The financial strength of an organization has a major impact on software product family development and management. Some of the financial indicators generally used in monitoring the performance of the business are; current ratio; debt to equity; debt coverage ratio; sales growth; net profit margin; return on assets; return on investment; and payback period. A successful software product line plays a key role in achieving the desired financial objectives of an organization. Some of the financial indicators such as current ratio, debt to equity and debt coverage ratio highlight the ability of an organization to invest in the software product line. Sales growth and net profit margin depict how successfully the software product line contributes to the business growth. Return on assets, return on investment and payback period indicates the potential of the software product line to achieve the long-term financial goals of an organization.

Asset management is a very important practice in the technical and financial planning of medium and long-term business ventures. It outlines action plans for the creation or acquisition, maintenance, operation, replacement and disposal of assets to provide an agreed-upon level of cost effective and sustainable development. Asset management involves deciding what assets to purchase, particularly in terms of the quantity and timing. Chen (2002) observes that computing asset management is a process or technology that helps manage computer hardware/software procurement and usage, facilitates license compliance, tracks inventory, enables change, or improves overall efficiency in software development. The mobilization of assets for effective use at the right time gives a tremendous advantage for an organization experiencing heavy competition in the market. The consolidated assets of an organization provide strategic strengths to the company, which helps in making long-range decisions to enter various segments of the market. Effective utilization of software assets is one of the major concerns for software development organizations. Such utilization has the potential to considerably reduce the development time, product defects, and cost of the software product. Software development organizations have shown a growing interest in the software product line concept because it deals with effective utilization of software assets.

## 3) Business Planning: Literature Review

Business planning spans strategic planning, business vision and innovation. Sutton (1993) describes strategic planning as a mechanism by which an organization collects and evaluates information about its own operations and its relationship to its environment, generates projections about future changes in that environment, and sets organizational goals based on those projections, which then serve as both a blueprint for change and a measure of progress. Strategic plans are the focus of an organization's endeavors to accomplish the desired level of achievement in a particular area. Strategic planning starts with elaborating strategic objectives. Harrison (1995) asserts that objectives indicate what management expects to accomplish, while planning sets forth how, when, where and by whom the objectives will be attained. Strategic planning is a continuous process within an organization; it determines business goals, evaluates the obstacles and defines approaches to deal with those obstacles. It outlines definite tasks for individuals, groups and for the entire organization, which are needed to accomplish these goals. Niemelä (2005) highlights eight different strategies for adopting software product lines in an organization: minimizing risk, extending market share, maximizing end-user satisfaction, balancing cost and potential, balancing cost, customer satisfaction and potential, and maximizing potential. Niemelä further concludes that a company has to evaluate the current status of their business, architecture, process, and organizational issues before making a decision about choosing one strategy out of those in order to achieve desired benefits.

The term "business vision" entails a description of where the organization stands several years in the future. Business vision is based on reality and on the current state of the organization, but it is focused on the future. It allows the organization to prepare action plans to introduce changes and improvements in current practices in order to reach future objectives. In practice, business vision is a statement that is prepared by senior management and is communicated to all members of the organization. The senior managers prepare the business vision after analyzing the organization's current situation and its impact on the external environment. The statement includes the identification of a desired future, and a well-established connection between the future and the present state. Overall, it serves as a link between one's experiences and knowledge of the past and present, with decisions about future. A successful business vision plan requires all the employees within the organization to participate and to clearly understand the vision statement. Wijnstra (2002) concludes that a complete business roadmap is needed to describe what is expected from the software product lines in the years to come and how it will fit in the plan for the release of new products.

Innovation is regarded as a by-product of research and development. Continuous research in attempting to understand a problem and discover possible solutions leads to innovation. Martensen and Dahlgaard (1999) maintain that an innovative strategy should be closely linked to the company's vision and overall business strategy, as well as being based on comprehensive and relevant information, both from inside the company and from the market. Innovation and continuous improvements in processes and products illustrate the capability of the organization to be creative and to be pioneers in product development. Organizations having intentions to capture a major share of the market in order to increase business spend heavy investments in research and development. Business objectives influence research and development efforts because the order of a product's entry into the market can make a significant difference in achieving strategic goals. Thus, research and development in technology, administration, processes and products, produces enduring results. Böckle (2005) highlights some measures of innovation management in software product line organizations, which include a planned innovation process, clear roles and responsibilities definition for innovation management structure. Böckle further stress that the evolution of the product portfolio, platform, variability model, and reference architecture shall be planned with further innovations in mind.

## B. Framework of Business Maturity Model

Most software process assessment frameworks such as CMM (Paulk *et al*., 1993), BOOTSTRAP (Kuvaja *et al*., 1994) consider staging while defining the maturity levels. The framework of business maturity model of software product line presented in this paper also uses the approach of staging. An organization at a particular maturity level must satisfy all business practices at that level. In their work, *van der* Linden *et al.* (2004) defines five business maturity scales for assessing the business dimension of software product line. These scales, in ascending order are: "reactive", "awareness", "extrapolate", "proactive" and "strategic". This work uses these business maturity scales to develop a framework consisting of a set of questionnaires for each maturity scale. The set of questions in the questionnaires are divided into eight business practices and cover the three business dimensions. The maturity level of an organization is determined from their extent of agreement of the organization with each question in the questionnaires. This is the first study of its kind within the area of software product lines to the best of our knowledge. All questionnaires are designed and written specifically for this business maturity model. Table-II illustrates the configuration of the framework of business maturity model. Each maturity level has set of questions and covers all the eight business practices used in this study. The number of questions varies for each maturity level as well as for each business practice. In the rest of this paper the following abbreviations for Market Orientation (MO), Relationships Management (RM), Order of Entry (OE), Financial Management (FM), Assets Management (AM), Strategic Planning (SP), Business Vision (BV) and Innovation (IN) are used. In the measuring instrument (questionnaires) of this maturity model the following symbols and abbreviations are used. The next sub-sections describe the characteristics of an organization dealing with software product line in terms of business maturity scales and the measuring instrument designed particularly for this business maturity model of software product line.

**BP.X.Y.Z** BP = Business Practice
X = Business Dimension (an integer)
Y = Maturity Level (an integer)
Z = Business Practice Number (an integer)

**Q.I.J.K.L** Q = Question
I = Business Dimension
J = Maturity Level (an integer)
K = Business Practice Number (an integer)
L = Question Number (an integer)

**Table-II: Framework of Business Maturity Model**

| Maturity Level | Business Practice & Number of Items in Assessment Questionnaire | | | | | | | | |
|---|---|---|---|---|---|---|---|---|---|
| | MO | RM | OE | FM | AM | SP | BV | IN | Total |
| Reactive | 1 | 2 | 2 | 2 | 2 | 1 | 1 | 1 | 12 |
| Awareness | 3 | 3 | 3 | 2 | 2 | 2 | 1 | 2 | 18 |
| Extrapolate | 3 | 2 | 2 | 3 | 3 | 3 | 3 | 3 | 22 |
| Proactive | 5 | 3 | 3 | 3 | 3 | 2 | 2 | 2 | 23 |
| Strategic | 3 | 1 | 2 | 2 | 3 | 2 | 3 | 2 | 18 |

### 1) Reactive (Level-1)

The first business maturity level of software product lines is level-1 “reactive”. The “reactive” stage of the business translates to organization that does not as yet have a stable and organized environment for software product line. There is no evidence that the organization performs business practices to establish coordination between business and software product line engineering activities. The organization tends to carry out multiple product development only as a reaction to market demands. There are no defined procedures for market surveys, customer profiling and product development schedules. There is a lack of strategic planning and absence of business vision in the organization. The financial health of the organization is not very convincing. A lack of understanding of software product line engineering methodology is present. The organization does not have the technological resources and skills to establish software product lines, although they have a growing interest in setting up a suitable infrastructure for product line engineering. Appendix-I illustrates the measuring instrument for assessing the software product line business maturity of an organization when it is a level-1 of “reactive”.

### 2) Awareness (Level-2)

The next business maturity level of software product lines is level-2, and is defined as “awareness”. The organizations at this level are generally aware of the potential benefits of software product lines. At the early stage of level-2, the organization is not able to align the business practices with product line engineering, except for their intensions to do so at some future stage. The marketing strategy of the organization starts providing feedback to product line engineering activities. The product development schedules are influenced by order of entry into the market. Organization starts managing software assets. The organization shows action and commitment to incorporate software product lines in its’ strategic plans and future direction. Organizational learning shows interest in the software product line concept. Overall the organization understands the importance of software product lines in achieving business goals. They are in the phase of establishing an infrastructure for software product lines. Lack of understanding of coordinated activities between business and engineering to launch software product line is present. Appendix-II illustrates the measuring instrument for assessing the software product line business maturity of an organization when it is a level-2 of “awareness”.

### 3) Extrapolate (Level-3)

An organization at level-3, which is also referred to as “extrapolate” is able to establish an infrastructure for software product lines. The organization is able to collect and disseminate market information. The organization makes software product line as a part of formal business planning. The scope of software product lines allows the organization to identify potential business cases. Market, customer and competitor orientations provide directions to the delivery schedules of software products. Strategic planning starts weighting software product lines a crucial entity to achieve business objectives. An initial set of activities required for establishing the infrastructure for software product lines are within the agenda of strategic planning. The organization understands the process methodology of software product lines and is able to start coordinating between business and product lines. There are efforts to align business and product line engineering results in innovative ways to capture a targeted market. The organization starts feeling the positive effect on their financial strength due to product line. Appendix-III illustrates the measuring instrument for the software product line business maturity of an organization when it is a level-3 of "extrapolate".

### 4) Proactive (Level-4)

The fourth level of business maturity of software product line is "proactive". An organization at this level has been able to establish coordination between business strategies and the software product line. The software product line decisions are influenced by business concerns. The software product line scope and product line requirements are aligned with the market. Product line delivery schedules accommodate market demands. The organization is able to maintain and update core asset repository. The organization has achieved the required skills and knowledge to launch and maintain the software product line. Strategic planning covers the product line requirements. The business vision of the organization foresees the importance of the product line in long-term business objectives. Innovative measures are introduced in product line engineering, which depicts the richness of organization culture in adopting product line engineering. The business decisions of the organization give weight to software product lines. Appendix-IV illustrates the measuring instrument for assessing the software product line business maturity of an organization when it is a level-4 of "proactive".

### 5) Strategic (Level-5)

The highest business maturity level is "strategic". An organization at level-5 considers software product family as a strategic asset, which can be mobilized to achieve desired business objectives. The market size of the product line has increased over a period of time and organization has established and maintained a position as a solution provider in the consumer market. The organization has sufficient resources and skills to give appropriate response to competitors' actions. The competitors consider the product line of the organization as a direct threat to their business. The organization exhibits the characteristics of early movers or even pioneers in product development. Software product line is contributing in improving the financial strength of the organization. The software product line plays an integral role in the business vision of the organization. The software product line plays a significant role in achieving the strategic objectives of the organization. Business is completely aligned with product line approach. The business decisions of the organization are strongly influenced by production plans of the product line. Appendix-V illustrates the measuring instrument for assessing the software product line business maturity of an organization when it is a level-5 of "strategic".

**Table-III: Performance Scale**

| Scale # | Linguistic Expression of Performance Scale | Linguistic Expression of BOOTSTRAP | Rating Threshold (%) | Value |
|---|---|---|---|---|
| 5 | Complete Agree | Complete Satisfied | $\geq$ 80 | 4 |
| 4 | Largely Agree | Largely Satisfied | 66.7 - 79.9 | 3 |
| 3 | Partially Agree | Partially Satisfied | 33.3 - 66.6 | 2 |
| 2 | Not Agree | Absent / Poor | $\leq$ 33.2 | 1 |
| 1 | Doesn't Apply | Doesn't Apply | - | 4 |

## C. Performance Scale

The maturity level of an organization is determined by measuring the ability of an organization to perform their business practices. Five levels scale is used to obtain performance rating. An ordinal rating "Completely Agree (4)", "Largely Agree (3)", "Partially Agree (2)" and "Not Agree (1)" as described in Table-III is used to measure each business practice. The rating threshold provides a set of quantitative measurements. These ratings reflect the agreement of the organization with each statement in the questionnaires. The scale point of 1 in Table-III "Doesn't Apply" is designed to increase the flexibility of the model and it is treated as equivalent to a value of 4 in the rating algorithm. It is important to note here that the performance scales and their rating threshold are kept close to the BOOTSTRAP methodology as illustrated in Table-III. We intentionally defined our performance scale in the previously existing approach in order to keep software product line business assessment close to existing popular scales, which were already in use, and have been validated and widely accepted in software industry. The rating thresholds values of the performance scales are also similar to BOOTSTRAP. We introduce some changes in the linguistic expressions of the performance scales. The major reason for these changes relates to the current design of the questionnaires of this study. Specifically, our questionnaires take self-assessment approach into account, where an organization is able to evaluate their business maturity by expressing their own level of agreement with the statements.

## D. Rating Method

The rating method adopted in this business maturity model for software product lines derives its foundations partially from BOOTSTRAP algorithm (Wang and King, 2000) of software process assessment. However, the structure of the rating method used different terminologies such as Performance Rating ($PR_{BP}$), Number of Agreed Statements ($NA_{BP}$), Pass Threshold ($PT_{BP}$), and Business Maturity Level (BML), discussed in detail as follows:

Let $PR_{BP}$ [I, J] be a rating of $I^{th}$ business practice of the $J^{th}$ maturity level. Then using the performance scale defined in Table-III, $PR_{BP}$ [I, J] can be rated as:

$PR_{BP}$ [I, J] = 4, if the extend of agreement with the statement is at least 80%.
= 3, if the extend of agreement with the statement is between 66.7-79.9%.
= 2, if the extend of agreement with the statement is between 33.3-66.6%.
= 1, if the extend of agreement with the statement is less than or equal to 33.2%.
= 4, if the statement does not apply in this assessment.

An $I^{th}$ statement at $J^{th}$ maturity level is considered agreed if $PR_{BP} [I, J] \geq 3$. If the number of statements agreed at maturity level "J" is $NA_{BP}$ [J] then it is defined by the expression:

$NA_{BP}$ [J] = Number of {$PR_{BP}$ [I, J] | Agreed}
= Number of {$PR_{BP} [I, J] \mid PR_{BP} [I, J] \geq 3$}

Table-IV illustrates the pass threshold of 80% at each maturity level; values are calculated to the nearest hundred. The maturity level is considered as pass or achieved if 80% of the statements in the questionnaire are agreed. If $N_{BP}$ [J] is the total number of statements at the $J^{th}$ maturity level then the pass threshold ($PT_{BP}$) at $J^{th}$ maturity level is defined as:

$PT_{BP} [J] = N_{BP} [J] * 80\%$

The organizational Business Maturity Level (BML) is defined as the highest maturity level at which the number of statements agreed is more than or equal to pass threshold ($PT_{BP}$ [J]), given by:

$BML = \max \{J \mid NA_{BP} [J] \geq PT_{BP} [J]\}$

Table-IV: Rating Threshold

| Maturity Level | Total Questions | Pass Threshold 80% |
|---|---|---|
| Reactive | 12 | 10 |
| Awareness | 18 | 14 |
| Extrapolate | 22 | 18 |
| Proactive | 23 | 18 |
| Strategic | 18 | 14 |

# III. RELIABILITY AND VALIDITY ANALYSIS OF QUESTIONNAIRES

The two most important aspects of precision in questionnaire-based assessments are reliability and validity. Reliability refers to the reproducibility of a measurement, whereas validity refers to the agreement between the value of a measurement and its true value. We conducted a pilot study and requested some organizations to provide us their extent of agreement with each statement in the questionnaires. This pilot study allows us to analyze the reliability and construct validity of the questionnaires designed in this study. The reliability of the questionnaires designed for five maturity levels were evaluated by using internal-consistency analysis method. Internal-consistency analysis was performed using coefficient alpha (Cronbach, 1951). Table-V reports the results of reliability analysis, the coefficient alpha ranges from 0.60 to 0.92. (Nunnally and Bernste, 1994) found that a reliability coefficient of 0.70 or higher for a measuring instrument was satisfactory. The other reliability literature such as (*van de* Ven and Ferry, 1980) suggests that a reliability coefficient of 0.55 or higher was satisfactory, and (Osterhof, 2001) concluded that 0.60 or higher is satisfactory. Our analysis shows that most of the questionnaire items developed for this business maturity model are satisfied with the criteria of (Nunnally and Bernste, 1994), whereas some of the items have alpha less than 0.70 but they still fall within the acceptable ranges of (*van de* Ven and Ferry, 1980) and (Osterhof, 2001). Our analysis shows that all the items are considered reliable as each construct has an alpha of 0.55 or higher therefore satisfying the acceptable ranges of alpha.

**Table-V: Reliability Analysis of Business Practices**

| Maturity Level | Business Practices | | | | | | | |
|---|---|---|---|---|---|---|---|---|
| | MO | RM | OE | FM | AM | SP | BV | IN |
| Reactive | * | 0.68 | 0.68 | 0.73 | 0.79 | * | * | * |
| Awareness | 0.82 | 0.92 | 0.87 | 0.76 | 0.90 | 0.90 | * | 0.82 |
| Extrapolate | 0.90 | 0.88 | 0.78 | 0.84 | 0.89 | 0.81 | 0.87 | 0.91 |
| Proactive | 0.91 | 0.85 | 0.66 | 0.84 | 0.69 | 0.72 | 0.55 | 0.72 |
| Strategic | 0.68 | * | 0.65 | 0.60 | 0.68 | 0.83 | 0.69 | 0.78 |

*** Construct has only one item evaluation of coefficient alpha is not possible**

Construct validity, according to Campbell and Fiske (1959), occurs when the scale items in a given construct move in the same direction, and, thus, are highly correlated. A principal component analysis (Comrey and Lee, 1992) performed and noted for all eight-business practices in each maturity level. Table-VI provides a measure of construct validity. We used eigen values (Kaiser, 1970) and scree plots (Cattell, 1966) as reference points to observe construct validity using principal component analysis. We used eigen value-one criterion, also known as Kaiser criterion (Kaiser, 1960; Stevens, 1986), which means any component having an eigen value greater than one is retained. Eigen values analysis reveals that the items present in questionnaires completely formed a single factor. The scree plots clearly showed a cut-off at the first component. Therefore, the construct validity can be regarded as being sufficient. It is important to note here that both principal component analysis and coefficient alpha requires more than one items in a construct to calculate eigen value and cronbach alpha. In the constructs of "MO","SP","BV" and "IN" at level-1 and "BV" and "RM" at level-2 and 5 respectively single items are present. Therefore we are unable to evaluate the reliability and validity analysis for those constructs and it is highlighted with "*" in Table-V. The presence of single items at level-1 is by design whereas in cases of level-2 and level-5 they were originally planned as two-item-construct, but did not proof reliable. This is one of the potential limitations of this work and further highlighted in the (Section IV, Sub-section "B").

**Table-VI: Construct Validity of Business Practices**

| Maturity Level | Business Practices | | | | | | | |
|---|---|---|---|---|---|---|---|---|
| | MO | RM | OE | FM | AM | SP | BV | IN |
| Reactive | * | 1.52 | 1.52 | 1.67 | 1.74 | * | * | * |
| Awareness | 2.25 | 2.63 | 2.40 | 1.63 | 1.85 | 1.83 | * | 1.70 |
| Extrapolate | 2.52 | 1.80 | 1.73 | 2.41 | 2.48 | 2.19 | 2.53 | 2.54 |
| Proactive | 3.76 | 2.32 | 1.86 | 2.35 | 2.01 | 1.61 | 1.53 | 1.56 |
| Strategic | 1.86 | * | 1.49 | 1.43 | 1.60 | 1.85 | 2.10 | 1.82 |

*** Construct has only one item PCA is not possible**

According to Campbell and Fiske (1959) convergent validity is the degree to which concepts should be related theoretically are interrelated in reality whereas discriminant validity is the degree to which concepts that should not be related theoretically are, in fact, not interrelated in reality. Campbell and Fiske (1959) propose Multitrait-Multimethod Matrix (MTMM) as a method to assess convergent and discriminant validity. The MTMM is a matrix of correlations among constructs to facilitate the interpretation of the assessment of convergent and discriminant validity. A high correlation among constructs leads to convergent validity whereas a low degree of correlation illustrates the discriminant validity. Table-VII illustrates the average inter-item correlation within construct to assess the convergent and discriminant validity. Face validity defines the extent to which the contents of a test or procedure look like they are measuring what they are supposed to measure. Face validity is a qualitative assessment of the measuring instrument. One of the approaches to carry out face validity is to get the feedback from the experts in the domain of the measuring instrument. We requested some of the experts actively involved in the process of software product line engineering to provide us their feedback about the face validity of the measuring instrument. After receiving their feedback we made some modifications in the structure and contents of the measuring instrument. The measurements of reliability and validity analysis show that the measurement procedures used in this business maturity model overall have the acceptable level of psychometric properties.

**Table-VII Convergent and Discriminant Validity Analysis**

| | Reactive | Awareness | Extrapolate | Proactive | Strategic |
|---|---|---|---|---|---|
| Reactive | 0.64 | | | | |
| Awareness | 0.21 | 0.48 | | | |
| Extrapolate | 0.10 | 0.05 | 0.73 | | |
| Proactive | 0.02 | 0.11 | 0.09 | 0.78 | |
| Strategic | 0.22 | 0.25 | 0.15 | 0.10 | 0.62 |

# IV. CASE STUDIES

In order to evaluate the business maturity level, we applied our model to two organizations currently involved in the process of software product line engineering. In order to protect the privacy of the two organizations, they will be referred to as "A" and "B". Organization "A" is involved in the business of telecommunication and is one of the biggest organizations in the mobile phone industry. Organization "B" is a software development firm that has been in business for over a decade and has software development sites all across the globe. Table VIII shows detailed assessment results of organization "A". The numerical values entered in each cell of Table VIII represent the organization's agreement with the statements in the questionnaires of each maturity level. Table-IX reports the summary of assessment results. It is important to note that according to the rating method discussed in Section II, Sub-Section D, a statement is considered agreed upon if the performance rating shown in Table-III is either greater than or equal to 3. Organization "A" is at the "Extrapolate" maturity level, which is level-3, while Organization “B” is at level-4, or the “Proactive” level. The following section elaborates on the assessment methodology used in this study.

## A. Assessment Methodology

- The two participating organizations are from North America. In terms of size, they are considered to be large since each has a total of over 3000 employees working in various departments.
- In the first stage of the study, we established contact with individuals in the two organizations to request their participation in this study. Specifically, we sent personalized emails to the individuals describing the scope and objectives of the study. Since the individuals contacted were working in the area of software product line engineering, their answers were directly applicable to the study. Furthermore, we informed the participants that the assessment being conducted was a part of a Ph.D. research project and that neither the identity of an individual nor of an organization would be disclosed in the resulting Ph.D. thesis or in any subsequent research publications.
- The questionnaires designed for this maturity model are used to measure the maturity of business dimension of each company’s software product line engineering. The individuals participating in the study were requested to provide the extent their agreement with each statement by using the performance scale ranging from 1 to 5. This performance scale is illustrated in Table III.
- Our assessment methodology uses a top down approach, where the more emphasized characteristics can be identified by moving from a lower to a higher level in the questionnaire. Consequently, the respondents complete the questionnaire by starting at Level 1 and finishing at Level 5.
- All of the participants in this study were volunteers, and no compensation of any form was offered or paid. We also told the respondents that if for any reason they did not want to answer any question, to please leave it blank.
- On average, the respondents to this study had been associated with their respective organizations for the last three years. The minimum educational qualification of the respondents was an undergraduate university degree and the maximum was a Ph.D. degree. Most of the respondents generally belonged to middle or senior technical management and were associated with the software development process. However, some of them were from other departments such as marketing, sales and business development. Some of the participants had policy making roles or were involved in implementing organizational strategies from the top to the bottom.
- We highlighted some major sources of data, such as documents, plans, models and actors, for the participants in the study. This was done to reduce the likelihood of inaccurate estimations from the respondents’ and to increase the reliability of the approach.
- Our assessment was not conducted by the usual on-site method. Specifically, we neither visited the organizations in person nor had meetings in person with the individual respondents to discuss the questionnaires. Our major source of contact and communication with the respondents was via email.
- We received multiple responses from each organization and thus limited the amount of bias in the sample. A variety of respondents from each organization provide a more accurate overall description of the company.

**Table-VIII Details of Assessment Result of Case Study “A”**

| Reactive Level-1 | | Awareness Level-2 | | Extrapolate Level-3 | | Proactive Level-4 | | Strategic Level-5 | |
|---|---|---|---|---|---|---|---|---|---|
| Question # | Value | Question # | Value | Question # | Value | Question # | Value | Question # | Value |
| Q 1.1.1.1 | 1 | Q 1.2.1.1 | 3 | Q 1.3.1.1 | 4 | Q 1.4.1.1 | 2 | Q 1.5.1.1 | 2 |
| Q 1.1.2.1 | 1 | Q 1.2.1.2 | 3 | Q 1.3.1.2 | 4 | Q 1.4.1.2 | 2 | Q 1.5.1.2 | 2 |
| Q 1.1.2.2 | 1 | Q 1.2.1.3 | 2 | Q 1.3.1.3 | 4 | Q 1.4.1.3 | 2 | Q 1.5.1.3 | 2 |
| Q 1.1.3.1 | 1 | Q 1.2.2.1 | 3 | Q 1.3.2.1 | 4 | Q 1.4.1.4 | 3 | Q 1.5.2.1 | 2 |
| Q 1.1.3.2 | 1 | Q 1.2.2.2 | 3 | Q 1.3.2.2 | 4 | Q 1.4.1.5 | 3 | Q 1.5.3.1 | 2 |
| Q 2.1.4.1 | 1 | Q 1.2.2.3 | 3 | Q 1.3.3.1 | 3 | Q 1.4.2.1 | 2 | Q 1.5.3.2 | 3 |
| Q 2.1.4.2 | 1 | Q 1.2.3.1 | 3 | Q 1.3.3.2 | 3 | Q 1.4.2.2 | 3 | Q 2.5.4.1 | 2 |
| Q 2.1.5.1 | 1 | Q 1.2.3.2 | 3 | Q 2.3.4.1 | 3 | Q 1.4.2.3 | 2 | Q 2.5.4.2 | 3 |
| Q 2.1.5.2 | 1 | Q 1.2.3.3 | 2 | Q 2.3.4.2 | 3 | Q 1.4.3.1 | 2 | Q 2.5.5.1 | 2 |
| Q 3.1.6.1 | 1 | Q 2.2.4.1 | 3 | Q 2.3.4.3 | 4 | Q 1.4.3.2 | 2 | Q 2.5.5.2 | 2 |
| Q 3.1.7.1 | 1 | Q 2.2.4.2 | 3 | Q 2.3.5.1 | 4 | Q 1.4.3.3 | 3 | Q 2.5.5.3 | 2 |
| Q 3.1.8.1 | 1 | Q 2.2.5.1 | 4 | Q 2.3.5.2 | 4 | Q 2.4.4.1 | 2 | Q 3.5.6.1 | 3 |
| | | Q 2.2.5.2 | 3 | Q 2.3.5.3 | 4 | Q 2.4.4.2 | 3 | Q 3.5.6.2 | 2 |
| | | Q 3.2.6.1 | 3 | Q 3.3.6.1 | 4 | Q 2.4.4.3 | 3 | Q 3.5.7.1 | 2 |
| | | Q 3.2.6.2 | 3 | Q 3.3.6.2 | 4 | Q 2.4.5.1 | 3 | Q 3.5.7.2 | 2 |
| | | Q 3.2.7.1 | 3 | Q 3.3.6.3 | 3 | Q 2.4.5.2 | 3 | Q 3.5.7.3 | 3 |
| | | Q 3.2.8.1 | 3 | Q 3.3.7.1 | 4 | Q 2.4.5.3 | 2 | Q 3.5.8.1 | 2 |
| | | Q 3.2.8.2 | 3 | Q 3.3.7.2 | 4 | Q 3.4.6.1 | 3 | Q 3.5.8.2 | 2 |
| | | | | Q 3.3.7.3 | 4 | Q 3.4.6.2 | 2 | | |
| | | | | Q 3.3.8.1 | 2 | Q 3.4.7.1 | 2 | | |
| | | | | Q 3.3.8.2 | 2 | Q 3.4.7.2 | 3 | | |
| | | | | Q 3.3.8.3 | 2 | Q 3.4.8.1 | 2 | | |
| | | | | | | Q 3.4.8.2 | 2 | | |

**Table-IX Summary of Assessment Results of Case Studies**

| Maturity Level | Total Questions | Pass Threshold 80% | Organization "A" $NA_{BP}$ | Organization "B" $NA_{BP}$ |
|---|---|---|---|---|
| Reactive | 12 | 10 | 0 | 0 |
| Awareness | 18 | 14 | 16 | 0 |
| Extrapolate | 22 | 18 | 19 | 22 |
| Proactive | 23 | 18 | 10 | 21 |
| Strategic | 18 | 14 | 4 | 9 |

* $NA_{BP}$ = Total number of agreed statements

## B. Limitations of the Assessment Methodology

Questionnaire-based maturity models are susceptible to certain limitations, which is the case with this model. Some of the limitations associated with this model of software product line engineering are as follows:

- The first limitation involves the degree of completeness of the model. Although we used eight different business factors, which were spread over five maturity levels, there may have been be other factors that influence the business process of software product lines. Other such contributing factors not considered in this model include organization size, economic conditions and political conditions.
- The second limitation of the methodology was the issue of subjective assessment. We used statistical techniques that were most commonly used in software engineering to ensure the reliability and validity of the questionnaire-based assessment approaches. However, our measurements were still largely based on the subjective assessment of individuals.
- Although we used multiple respondents within the same organization to reduce bias, bias is still a core issue in decision-making and evaluating questionnaire-based responses. Product line engineering is a relatively new concept in software development, and not many of the organizations in the software industry have institutionalized and launched this concept. Hence, collecting data for determining the level of reliability and validity of the various assessment items from the software industry was a limitation.
- The degree of respondent participation also affected the accuracy of the results. As previously mentioned, we asked the respondents to consult major sources of relevant data in their organizations to reduce the possibility of inaccurate judgment when filling in questionnaires. However, the data collection was largely dependent on the individuals' efforts to obtain the required information before responding to the statements presented in the questionnaire.
- Our assessment methodology did not account for the role of independent assessors even though their role is an important aspect of maturity assessment modeling. Their role defines the level of coordination between the assessor and the internal assessment team and provides for an evaluation. The current case studies are based on self-assessment.

- The methodology evaluates and quantifies the maturity level of the different business factors as well as gauges the overall maturity level of the business dimension. However, our maturity model does not provide any guidelines for an improvement process, which we consider to be a subsequent project emanating from this study.
- There are six single item constructs in the measuring instruments for which reliability and validity analysis were not performed. This we consider as one of the limitations of the assessment methodology proposed in this work.

Although the business maturity model presented in this paper has both some general and specific limitations, it still provides a comprehensive approach to evaluating the maturity level of the business dimension for software product line engineering. Furthermore, it provides a suitable foundation for future research in this area.

### C. Utilization of the Business Maturity Assessment Model

One of the advantages of using maturity models in software engineering is that they have the ability to obtain inside information about the current maturity level of the different process-related activities in a particular organization. Ideally, this information provides a basis for improvement plans and activities. Furthermore, maturity models are also advantageous to individual organizations because companies with high maturity ratings are more attractive to potential customers. We summarized the advantages of the business maturity model from different perspectives as: software engineering research, various organizational aspects, product development and process improvements.

- Overall, the maturity level model presented in this work provides information that can be used to improve the organization’s process methodology and complementary product development activities within the organization.
- The overall business performance of an organization depends on a number of critical factors. Since technologies and business opportunities are evolving rapidly, companies must monitor the factors affecting the business performance. The monitoring of business factors helps in achieving the company’s ultimate goal of developing and profiting from products. The maturity assessment model presented in this work helps companies to monitor and evaluate their overall business process.
- The model presented also highlights a methodology for evaluating some of the business factors in a company. This evaluation provides inside information about the factors that can be improved upon by management. For example, if management discovered that market orientation is at a lower maturity level then they could introduce changes to the marketing strategy and plans to improve it. Such improvements could subsequently help in the product development process, which is the ultimate goal of the organization.
- The software product line is gaining popularity and many organizations around the world are currently involved in applying this concept. Our model provides an early conceptual framework for the maturity assessment of software product line engineering. Consequently, this area of study still requires future contributions from software engineering researchers.

## V. FINAL REMARKS

The overall engineering efforts of software product line development and management have been divided into four dimensions of business, architecture, software engineering process and organizational aspects. The software product line process assessment is relatively an area where only conceptual work has been done yet. This papers’ main contribution is a methodology to evaluate the business dimension of software product line engineering and enhances further understanding of the business aspects of software product line. The framework of the model consists of assessment questionnaires for the five levels maturity scale, performance scales and a rating method. Thus, this research contributes towards establishing a comprehensive and unified strategy for process maturity evaluation of software product line. The case studies conducted in this research showed the performance of the two organizations in their business dimension of software product line. This research also reinforces current perceptions that the software product line requires a comprehensive alignment of inter-disciplinary business strategies with software engineering activities. Currently, we are working on the development of a business evaluation tool to automate the assessment process of the model. Besides its general and specific limitations, the business maturity model presented in this paper contributes significantly in the area of software product line by addressing a topic of immense importance.

# Appendix-I

BP.1.1.1 Market orientation

- Q.1.1.1.1 The organization has not yet acquired adequate knowledge and skills to gather information about the market.

BP.1.1.2 Relationships management

- Q.1.1.2.1 The organization is not able to attract new and retain existing customers.
- Q.1.1.2.2 The organization has complex business processes and customers are not satisfied with them.

BP.1.1.3 Order of entry to the market

- Q.1.1.3.1 Order of entry to the market is not an issue of concern in product development schedules.
- Q.1.1.3.2 The organization does not conduct market reviews to update product launch timing.

BP.2.1.4 Financial management

- Q.2.1.4.1 The organization is not able to reduce its debt.
- Q.2.1.4.2 The sales do not grow over a period of time.

BP.2.1.5 Assets management

- Q.2.1.5.1 The organization does not have formal plans to maintain and mobilize its assets.
- Q.2.1.5.1 Software assets are occasionally used in new product development.

BP.3.1.6 Strategic planning

- Q.3.1.6.1 Software product line is not a part of strategic plan of the organization.

BP.3.1.7 Business vision

- Q.3.1.7.1 The employees have no idea where the organization is going in next ten years.

BP.3.1.8 Innovation

- Q.3.1.8.1 The organization has no research and development setup.

# Appendix-II

BP.1.2.1 Market orientation

- Q.1.2.1.1 Market intelligence is shared, but there is a lack of defined inter-communication protocol among external and internal entities of the organization.
- Q.1.2.1.2 The organization occasionally collects and analyzes data from the consumer market to identify opportunities for new product development.
- Q.1.2.1.3 The marketing plans of the organization are influenced by actions of competitors.

BP.1.2.2 Relationships management

- Q.1.2.2.1 The organization gives consideration to the complaints of the customers' and is able to resolve their issues.
- Q.1.2.2.2 Organization listen the requirements of the customers in making decisions about new products.
- Q.1.2.2.3 The organization is able to reduce the number of complaints from customers over the period of time.

BP.1.2.3 Order of entry to the market

- Q.1.2.3.1 The organization gives weight to the order of entry to the market and product launch timing influences development schedules.
- Q.1.2.3.2 The organization generally develops products in response to the competitor actions.
- Q.1.2.3.3 The organization occasionally conducts market reviews and updates the development and delivery schedule of the software product line.

BP.2.2.4 Financial management

- Q.2.2.4.1 The organization is able to maintain its debt.
- Q.2.2.4.2 There is no change in the net profit margin during the last two years.

BP.2.2.5 Assets management

- Q.2.2.5.1 The organization has established a policy of managing assets for the software product line.
- Q.2.2.5.2 The software product line assets information is collected at need to know basis by most of the personnel involved in product development.

BP.3.2.6 Strategic planning

- Q.3.2.6.1 Software product line is considered as an option in the strategic plans of the organization.
- Q.3.2.6.2 The strategic planning identifies key market segments for the software product line.

BP.3.2.7 Business vision

- Q.3.2.7.1 The organization is in the planning phase of setting up future goals and positioning

software product line as an important tool to achieve desired goals.

BP.3.2.8 Innovation
- Q.3.2.8.1 The organization has established an infrastructure for research and development in software product lines.
- Q.3.2.8.2 The employees have opportunities to participate in problem solving and idea generation activities for the software product line.

# Appendix-III

BP.1.3.1 Market orientation
- Q.1.3.1.1 The organization has an established defined inter-communication protocol among external and internal entities for the dissemination of market intelligence.
- Q.1.3.1.2 The domain engineering activity of product line engineering identifies the potential market segment.
- Q.1.3.1.3 The organization uses feedback from customers' to develop new products or services.

BP.1.3.2 Relationships management
- Q.1.3.2.1 The organization has a well-established system to quickly extract, manipulate and produce data for profitability analysis, customer profiling, and retention modeling.
- Q.1.3.2.2 The organization simplifies business processes regularly to enhance customer experience and satisfaction.

BP.1.3.3 Order of entry to the market
- Q.1.3.3.1 The products developed from the software product line enter into the market at appropriate time.
- Q.1.3.3.2 The organization regularly conducts market reviews and updates the development and delivery schedule of the software product line.

BP.2.3.4 Financial management
- Q.2.3.4.1 The organization is able to reduce its debt.
- Q.2.3.4.2 The sales grow over a period of time.
- Q.2.3.4.3 The return on assets increases over a period of time.

BP.2.3.5 Assets management
- Q.2.3.5.1 The organization has allocated sufficient resources for managing core assets of software product line.
- Q.2.3.5.2 The assets within the software product line repository are consistent with the scope of the software product line.
- Q.2.3.5.3 The software product line assets information is well communicated to all personnel involved in the product development.

BP.3.3.6 Strategic planning
- Q.3.3.6.1 The strategic planning allocates resources for software product line development.
- Q.3.3.6.2 The strategic plans define how an organization will achieve the technological capability to successfully adopt the concept of the software product line.
- Q.3.3.6.3 The strategic plans portray what should be developed from the software product line.

BP.3.3.7 Business vision
- Q.3.3.7.1 The organization has a well-documented business vision statement.
- Q.3.3.7.2 In the business vision of the organization software product line aims at retaining current customers and attracting future ones.
- Q.3.3.7.3 The software product line is considered essential for the organization to reach its future goals.

BP.3.3.8 Innovation
- Q.3.3.8.1 The organization has defined a road map for research and development in software product lines.
- Q.3.3.8.2 The organization allocates resources to research and development in the software product line.
- Q.3.3.8.3 The innovations in the software product line are aligned with the existing business goals

of the organization.

# Appendix-IV

BP.1.4.1 Market orientation

- Q.1.4.1.1 The organization uses feedback from customers to the improve quality of products and services.
- Q.1.4.1.2 The organization has adequate resources and skills to gather information about the market.
- Q.1.4.1.3 The organizations regularly collect and analyze data from the consumer and market to identify opportunities for new market segments.
- Q.1.4.1.4 The scope of software product line covers the customers' requirements.
- Q.1.4.1.5 The organization explicitly considers competitors as their top priority while developing market plans.

BP.1.4.2 Relationships management

- Q.1.4.2.1 The organization attracts new and existing customers through personalized communication and innovative targeting methods.
- Q.1.4.2.2 The organization has an established promotions strategy to attract new customers and retain existing ones.
- Q.1.4.2.3 The organization is able to retain customers over a long period of time.

BP.1.4.3 Order of entry to the market

- Q.1.4.3.1 The organization has the potential of being first in the market to introduce new products.
- Q.1.4.3.2 The delivery schedule of products out of software product line helps in increasing the market presence of the organization.
- Q.1.4.3.3 The software product line is able to meet the demands of the delivery schedule requirements of the customers.

BP.2.4.4 Financial management

- Q.2.4.4.1 The net profits margin increase over a period of time.
- Q.2.4.4.2 The payback period decrease over a period of time.
- Q.2.4.4.3 The software product line fits in the financial model of the organization.

BP.2.4.5 Assets management

- Q.2.4.5.1 The assets of the software product line are dynamic, and continuously grow as the production proceeds.
- Q.2.4.5.2 The software assets have significantly reduced the development cycle of the software product line.
- Q.2.4.5.3 The software assets are consistent with the production constraints and the production plan of the software product line.

BP.3.4.6 Strategic planning

- Q.3.4.6.1 The software product line is aligned with the strategic plans of the organization.
- Q.3.4.6.2 The strategic plans highlight an evolution in the software product line under changing business conditions.

BP.3.4.7 Business vision

- Q.3.4.7.1 The business vision is communicated to all members of the organizations and they are committed to achieve organizational goals.
- Q.3.4.7.2 Software product line is a part of the business vision of the organization.

BP.3.4.8 Innovation

- Q.3.4.8.1 The organizational culture supports innovation in the software product line.
- Q.3.4.8.2 The management supports reactive and proactive innovations in the software product line process.

# Appendix-V

BP.1.5.1 Market orientation

- Q.1.5.1.1 The organization successfully responds to the actions of competitors and is able to decrease the number of competitors over the period of time.
- Q.1.5.1.2 The scope of software product line is aligned with the requirements of the targeted market.
- Q.1.5.1.3 The organization is able to increase market size, and there is a steady increase in market growth over time.

BP.1.5.2 Relationships management

- Q.1.5.2.1 The organization is able to establish a balance in customer and product-centered approaches in product development.

BP.1.5.3 Order of entry to the market

- Q.1.5.3.1 The organization is regarded as pioneer in product development.
- Q.1.5.3.2 The customers are satisfied with the product launch timing of the software product line.

BP.2.5.4 Financial management

- Q.2.5.4.1 The return on investment increases over a period of time.
- Q.2.5.4.2 The software product line is contributing in strengthening the financial position of the organization.

BP.2.5.5 Assets management

- Q.2.5.5.1 The organization maintains information about assets, as well as their versions and utilization history in product development.
- Q.2.5.5.2 The assets management of the organization is aligned with the strategic plans of management.
- Q.2.5.5.3 The software assets management satisfies the cost to benefits ratio for the organization.

BP.3.5.6 Strategic planning

- Q.3.5.6.1 The software product line plays a significant role in achieving the strategic objectives of the organization.
- Q.3.5.6.2 The organizational strategic planning place the software product lines an important strategic consideration and even a strategic asset.

BP.3.5.7 Business vision

- Q.3.5.7.1 The business vision is regularly reviewed, updated as needed and communicated to all in the organization.
- Q.3.5.7.2 The employees understand the importance of the software product line in the business vision and feel that the organization can realistically achieve their targets.
- Q.3.5.7.3 The software product line plays a significant role in achieving the business vision of the organization.

BP.3.5.8 Innovation

- Q.3.5.8.1 The organization has successfully employed innovations in the software product line development.
- Q.3.5.8.2 The organization is committed in learning and improving knowledge in the area of software product lines.

## Authors Biography

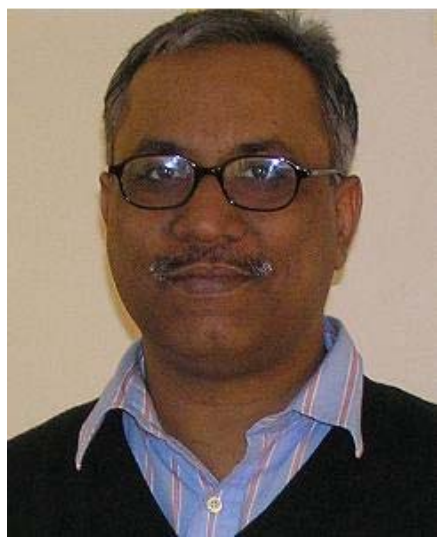

Faheem Ahmed received his MS (2004) and Ph.D. (2007) in Electrical Engineering from the University of Western Ontario, London, Canada. Currently he is assistant professor at College of Information Technology, UAE University, Al Ain, United Arab Emirates. Ahmed had many years of industrial experience holding various technical positions in software development organizations. During his professional career he has been actively involved in the life cycle of software development process including requirements management, system analysis and design, software development, testing, delivery and maintenance. Ahmed has authored and co-authored many peer-reviewed research articles in leading journals and conference proceedings in the area of software engineering. Ahmed's current research interests are Software Product Line, Software Process Modeling, Software Process Assessment, and Empirical Software Engineering. For more information please visit: http://faculty.uaeu.ac.ae/f_ahmed.

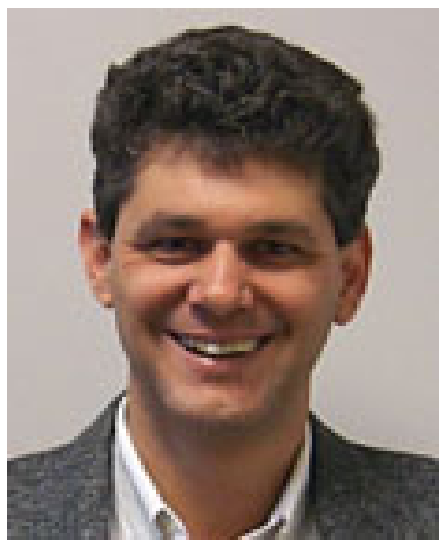

Luiz Fernando Capretz has almost 30 years of international experience in the software engineering field as a practitioner, manager and educator. Having worked in Brazil, Argentina, U.K., Japan, Italy and the United Arab Emirates, he is currently an Associate Professor and the Director of the Software Engineering Program at the University of Western Ontario, Canada. His present research interests include software engineering (SE), human factors in SE, software estimation, software product lines, and software engineering education. Dr. Capretz received his Ph.D. in Computing Science from the University of Newcastle upon Tyne (U.K.), his M.Sc. in Applied Computing from the National Institute for Space Research (INPE, Brazil), and his B.Sc. in Computer Science from State University of Campinas (UNICAMP, Brazil). Further information can be found at: http://www.eng.uwo.ca/people/lcapretz.